\documentclass[final,3p,times,twocolumn]{elsarticle}
 \biboptions{comma,sort&compress}
\usepackage{here}
 \usepackage{graphicx}
  \usepackage{epsfig}
   \usepackage{axodraw}
\def\nin{\noindent}
\def\beq{\begin{equation}}
\def\eeq{\end{equation}}
\def\bea{\begin{eqnarray}}
\def\eea{\end{eqnarray}}
\def\nnb{\nonumber}

\def\b{$\bullet~$}
\def\ga{\left(}
\def\dr{\right)}
\journal{Nuc. Phys. (Proc. Suppl.)}

\begin{document}

\begin{frontmatter}



\title{The $\sigma$ and $f_0(980)$ from $K_{e4}\oplus\pi\pi$, $\gamma\gamma$ scatterings, ~$J/\psi,\phi\to\gamma \sigma_B$ and $D_s\to l\nu \sigma_B$ }

 \author[label1]{G. Mennessier }
\ead{gerard.mennessier@lpta.univ-montp2.fr}

 \author[label1]{S. Narison\corref{cor1} }
   \address[label1]{Laboratoire
de Physique Th\'eorique et Astroparticules, CNRS-IN2P3,  
Case 070, Place Eug\`ene
Bataillon, 34095 - Montpellier Cedex 05, France.}
\cortext[cor1]{Corresponding author}
\ead{snarison@yahoo.fr}

 \author[label1,label3,label4]{X.-G. Wang\corref{cor2}}
  \address[label3]{Department of Physics, Peking University, Beijing 100871, China.}
\cortext[cor2]{China scholarship council fellow under contract n$^0$ 2009601139.}
\fntext[label4]{Speaker}
\ead{wangxuangong@pku.edu.cn}

%\cortext[cor2]{China Scholarship Council
%fellow under contract No.2009601139.} \ead{wangxuangong@pku.edu.cn}
%\author{}

%\address{}

\begin{abstract}
\noindent We extract the pole positions, hadronic and $\gamma\gamma$ widths of $\sigma$
and $f_0(980)$, from $\pi\pi$ and $\gamma\gamma$ scattering data
using an improved analytic K-matrix model.  Our results favour a large gluon component
for the $\sigma$ and a $\bar ss$ or/and gluon component for the $f_0(980)$ but neither a large four-quark nor a molecule component. Gluonium $\sigma_B$ production from $J/\psi,~\phi$ radiative and $D_s$ semi-leptonic decays are also discussed.

\end{abstract}

\begin{keyword}
$\gamma\gamma$ and $\pi\pi$ scatterings, radiative decays, light scalars, gluonia, four-quark states, QCD spectral sum rules.


\end{keyword}

\end{frontmatter}

% \linenumbers

\section{Introduction}
%\label{}
\nin
%%%%%%%%%%%%
 -The hadronic and $\gamma\gamma$ couplings of light scalar mesons could
 provide an important information about their nature.\\
 -K-matrix model has been used to describe $\pi\pi$ and $\gamma\gamma$
 processes~\cite{Mennessier83}.\\
 -This model is improved by introducing a form factor {\em shape function} in a single
 channel~\cite{Ochs08}, which is generalized \cite{KMN,MNW,MNW1} to the coupled channels case in order to extract the pole postions and the previous couplings of
 the $\sigma$ and $f_0(980)$. \\
 In this talk, we review these recent results.
%%%%%%%%%%%%
\section{Phenomenology of $\pi\pi$ scattering}
%\label{}
\nin
%%%%%%%%%%%%
\subsection*{\b\bf 1 channel $\oplus$ 1 "bare" resonance}
\nin
We introduce a real analytic form factor {\em shape
function}~\cite{Ochs08}:
\begin{equation}
f_p(s)=\frac{s-s_{AP}}{s-\sigma_{DP}},\ \ \ P=\pi,K
\end{equation}
It allows for an Alder zero $s=s_{AP}$ and a pole at
$\sigma_{DP}<0$ to simulate left hand singularities. The unitary
$I=0$ S wave $\pi\pi$ scattering amplitude is then written as:
\begin{equation}
T_{PP}=\frac{g_{P}^2 f_P(s)}{s_R-s-g_P^2 \tilde{f}_P(s)}=\frac{g_P^2
f_P(s)}{\mathcal {D}_P(s)},
\end{equation}
where:
\begin{equation}
\mathrm{Im}\mathcal{D}=\mathrm{Im}(-g_{\pi}^2\tilde{f}_P)=-(\theta\rho_P)g_{P}^2
f_P\ .
\end{equation}
and hence:
\begin{equation}
\mathrm{Im}(\tilde{f}_P)=(\theta\rho_P)f_P\ ,\ \ \ \
\rho_P=\sqrt{1-4m_{P}^2/s}.
\end{equation}
The real part of $\tilde{f}_P$ is obtained from a dispersion
relation with subtraction at $s=0$,
\begin{equation}
\tilde{f}_P(s)=\frac{2}{\pi}(h_0(s)-h_0(0))\ ,
\end{equation}
where $h_0(s)$ has been defined  in Ref.~\cite{Ochs08}.
%%%%%%%%%%%%%%%%%%%%%%%%%%%%%%%%%%%%%%%%
\subsection*{\b\bf 0 bare resonance$\equiv\lambda\phi^4$ model}
\begin{figure}[hbt]
\centerline{\includegraphics[width=4.2cm]{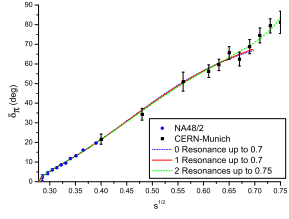}}
%{\epsfig{figure=mpsi2mc.eps,height=70mm}}
\caption{\scriptsize The fit result of $\pi\pi$ $I=0$ S-wave phase
shift.} \label{fig1}
\end{figure}

\nin
In this case, one can introduce another shape function $f_2(s)$:
\begin{equation}
T_{PP}=\frac{\Lambda f_{2}(s)}{1-\Lambda\tilde{f}_2(s)}\ ,\ \ \ \
f_2(s)=\frac{s-s_{AP}}{(s-\sigma_{D1})(s-\sigma_{D2})}\ .
\end{equation}
where $\sigma_{D1}=\sigma_{D_{\pi}}$, and
%\begin{equation}
$
\tilde{f}_2(s)=\frac{2}{\pi}[h_2(s)-h_2(0)]\ .
$
%\end{equation}
The single channel results are shown in Fig.~\ref{fig1} and
Table~\ref{tab:param1}. The result of $\lambda\phi^4$ model shows that the existence
of $\sigma$ is not an artifact of a "bare" resonance entering into the
parametrization of $T_{PP}$.
%%%%%%%%%%%%%%%%%%%%%%%
\nin
 %%%%%%%%%%%%%%%%%%%%%%%%%%%%%%%
{\scriptsize
\begin{table}[hbt]
\setlength{\tabcolsep}{0.35pc}
 \caption{\scriptsize    Values of the bare parameters in $\mathrm{GeV}^{d}(d=1,2)$.}
    {\footnotesize
\begin{tabular}{lcccl}\hline
Outputs & 0 res. & 1 res. & 2 res. & Average \\ \hline $s_A$&
0.009(6)&0.0094(fixed)&0.0094(fixed)& \ \\
$\sigma_{D\pi}$&6.2(3.2)&1.41(7)&1.78(10)& \ \\
$\sigma_{D2}$&7.6(4.5)&-&-&\ \\
$s_{Ra}$&-&1.94(9)&26.97(1.54)&\ \\
$\Lambda$&108(34)&-&-&\ \\
$g_{\pi a}$&-&2.54(8)&10.42(30)&\ \\
$s_{Rb}$&-&-&0.61(31)&\ \\
$g_{\pi b}$&-&-&-0.39(8)&\ \\
$\chi^2_{d.o.f}$&$\frac{12.04}{14}=0.86$&$\frac{11.73}{15}=0.78$&
$\frac{12.71}{16}=0.79$&\
\\
\ & \ & \ & \ & \ \\
$M_{\sigma}$&468(181)&456(19)&448(18)&452(13) \\
$\Gamma_{\sigma}/2$&261(211)&265(18)&260(19)&259(16) \\
$|g_{\sigma\pi^+\pi^-}|$&2.58(1.31)&2.72(16)&2.58(14)&2.64(10)
\\ \hline
\end{tabular}
} \label{tab:param1}
\end{table}
} \nin 
%%%%%%%%%%%%%%%%%%%%%%%%%%%%%%%%%%%%%%%%%%%%%%%%%%%%%%%%%%
\subsection*{\b\bf 2 channels $\oplus$ 2 "bare" resonances} 
\nin
The
generalization to coupled channels is conceptually straightforward. We
consider the $\pi\pi-K\bar{K}$ coupled channels and introduce 2 "bare"
resonances labeled $a$ and $b$, with bare masses squared $s_{Ra}$
and $s_{Rb}$. To leading order in SU(3) breakings, we shall approximately 
work in the {\em minimal case} with only one
shape function.
%\begin{equation}
%f(s)=\frac{s-s_A}{s+\sigma_D}\ .
%\end{equation}
The phase shifts and inelasticity $\eta$ are defined by:
\begin{equation}
\eta e^{2i\delta_P}=1+2{\rm i}~\rho_P T_{PP}\ .
\end{equation}
We refer to~\cite{MNW} for the explicit expressions of $T_{PP}$.
%%%%%%%%%%%%%%%%%%%%%%%%%%%%%%%%%%%%%%%%%%%%%%%%%%%%%%%%%%%%%%%%%
{\scriptsize
\begin{table}[hbt]
\setlength{\tabcolsep}{1.15pc}
 \caption{\scriptsize Different data used for each set.}
    {\footnotesize\begin{tabular}{cccc}\hline
Input & Set 1& Set 2& Set 3 \\ \hline
$\delta_{\pi}$ &
\cite{NA48}-\cite{CERN-Munich}& \cite{NA48}-\cite{CERN-Munich} &
\cite{NA48}-\cite{CERN-Munich} \\
$\eta$ & \cite{CERN-Munich}& \cite{Hyams}& \cite{Cohen80} \\
$\delta_{\pi K}=\delta_{\pi}+\delta_K$ & \cite{CERN-Munich} &
\cite{Cohen80} & \cite{Cohen80} \\ \hline
\end{tabular}
} \label{tab:data}
\end{table}
} \nin
%%%%%%%%%%%%%%%%%%%%%%%%%%%%%%%%%%%%%%%
 We analyze 3 cases using different groups of $\pi\pi-K\bar{K}$ data, which are
shown in Table~\ref{tab:data}. With these choices, we expect to
span all possible regions of space of parameters.
%%%%%%%%%%%%%%%%%%%%%%%%%%%%%%%%%%%
\begin{figure}[hbt]
\centerline{\includegraphics[width=4.2cm]{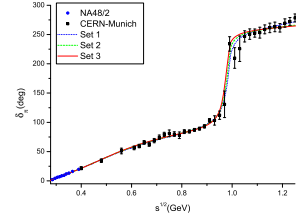}}
\centerline{\includegraphics[width=4.2cm]{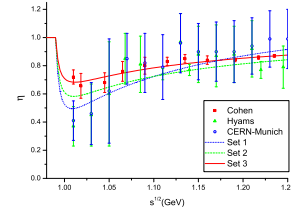}}
\centerline{\includegraphics[width=4.2cm]{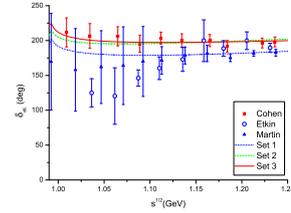}}
%{\epsfig{figure=mpsi2mc.eps,height=70mm}}
\caption{\scriptsize The coupled channels fit results.} \label{fig2}
\end{figure}
\nin
%%%%%%%%%%%%%%%%%%%%%%%%%%%%
{\scriptsize
\begin{table}[hbt]
\setlength{\tabcolsep}{0.5pc}
 \caption{\scriptsize    Values of the bare parameters in $\mathrm{GeV}^{d}(d=1,2)$.}
    {\footnotesize
\begin{tabular}{lccc}\hline
Outputs & Set 1 & Set 2 & Set 3 \\ \hline $s_A$ & 0.016$\pm$0.004 &
0.013$\pm$ 0.006 & 0.010$\pm$0.006 \\
$\sigma_D$ & 0.740$\pm$0.097 & 0.909$\pm$0.201 & 1.116$\pm$0.262 \\
\ & \ & \ & \ \\
$s_{Ra}$ & 4.112$\pm$0.499 & 2.230$\pm$0.271 & 2.447$\pm$0.298 \\
$g_{\pi a}$ & -0.557$\mp$0.177 & 0.864$\pm$0.391 & 0.997$\pm$0.516 \\
$g_{Ka}$ & 3.191$\pm$0.499 & 1.458$\pm$0.262 & 1.684$\pm$0.363 \\
\ & \ &\ &\ \\
$s_{Rb}$ & 1.291$\pm$0.062 & 1.187$\pm$0.094 & 1.354$\pm$0.149 \\
$g_{\pi b}$ & -1.562$\mp$0.117 & -1.527$\mp$0.134 & -1.756$\mp$0.183 \\
$g_{Kb}$ & 0.748$\pm$0.062 & 0.999$\pm$0.149 & 1.159$\pm$0.261 \\
\ & \ & \ & \ \\
$\chi^2_{d.o.f}$ & 70.6/77=0.914 & 48.8/64=0.759 & 44.3/58=0.763 \\
\hline
\end{tabular}
} \label{tab:param2}
\end{table}
} \nin
%%%%%%%%%%%%%%%%
{\scriptsize
\begin{table}[hbt]
\setlength{\tabcolsep}{0.55pc}
 \caption{\scriptsize Physical quantities. Mass and width are in unit of $\mathrm{MeV}$, while the couplings are in $\mathrm{GeV}$.}
    {\footnotesize
\begin{tabular}{lllll}\hline
Outputs & Set 1 & Set 2 & Set 3 & Average\\ \hline $M_{\sigma}$ &
435(74) & 452(72) & 457(76) & 448(43)\\
$\Gamma_{\sigma}/2$ & 271(92) & 266(65) & 263(72) & 266(43)\\
$|g_{\sigma\pi^+\pi^-}|$ & 2.72(78) & 2.74(61) & 2.73(61) &
2.73(38)\\
$|g_{\sigma K^+K^-}|$ & 1.83(86) &
0.80(55) & 0.99(68) & 1.06(38)\\
\ & \ & \ & \ & \ \\
$M_{f}$ & 989(80) & 982(47) & 976(60) & 981(34)\\
$\Gamma_{f}/2$ & 20(32) & 18(16) & 18(18) & 18(11)\\
$|g_{f\pi^+\pi^-}|$ & 1.33(72) & 1.22(60) & 1.12(31) & 1.17(26)\\
$|g_{fK^+K^-}|$ & 3.21(1.70) & 2.98(70) & 3.06(1.07) & 3.03(55)\\
\hline
\end{tabular}
} \label{tab:param3}
\end{table}
}  \nin
%%%%%%%%%%%%%%%%%%%%%%%%%%%%%%%%%%%%%%%%%%%%%%%%%%
The fit results are shown in Fig.~\ref{fig2} and
Table~\ref{tab:param2}, from which we derive the pole positions and
hadronic couplings of the $\sigma$ and $f_0(980)$.
%%%%%%%%%%%%%%%%%%%%%%%%%%%%%%%%%%%%%%%%%%%%%%%%%%%%%%%%
{\scriptsize
\begin{table}[hbt]
\setlength{\tabcolsep}{2pc}
 \caption{\scriptsize Mass and width in $\mathrm{MeV}$ of $\sigma$ on the complex plane.}
    {\footnotesize
\begin{tabular}{ll}\hline
Processes & $M_{\sigma}-i\Gamma_{\sigma}/2$  \\ \hline
%\ &\ &\ \\
%Our work & \ & \ \\
$K_{e4}\oplus\pi\pi\rightarrow\pi\pi$ & $452(13)-i259(16)$   \\
$K_{e4}\oplus\pi\pi/K\bar{K}$ & $448(43)-{\rm i}~266(43)$   \\
Average & $452(12)-{\rm i}~260(15)$ \\
%\\ 
\hline
\end{tabular}
} \label{tab:param4}
\end{table}
} \nin
%%%%%%%%%%%%%%%%%%%%%%%%%%%%%%%%%%%%%%%%%%%%%%%%%%%%%%%%%%%%%%%%%%%%%%%%%%
\subsection*{\bf \b Final results}
-- For the $\sigma$, we take the average value of the single and
coupled channels results:
\begin{eqnarray}
M_{\sigma}-{\rm i}\Gamma_{\sigma}/2&=&452(12)-{\rm i}~260(15)~\mathrm{MeV},\nonumber\\
|g_{\sigma\pi^+\pi^-}|&=&2.65(10)~\mathrm{GeV},\nonumber\\
r_{\sigma\pi K}&\equiv&\frac{|g_{\sigma
K^+K^-}|}{|g_{\sigma\pi^{+}\pi^{-}}|}=0.37(6)\ .
\end{eqnarray}
The mass and width shown in Table~\ref{tab:param4} are in good
agreement with other ones in the
literature.\\
The sizeable coupling of the $\sigma$ to $\bar{K}K$
disfavors the usual $\pi\pi$ molecule and 4-quark assignment of
 the $\sigma$, where this coupling is expected to be negligible. Moreover, a broad $\pi\pi$ width (compared with
$\rho$ meson) can not be explained within a
 $q\bar{q}$ scenario.\\
-- For the $f_0(980)$, we have:
\begin{eqnarray}
M_{f}&=&981(34)-{\rm i}~18(11)~\mathrm{MeV},\nonumber\\
|g_{f\pi^{+}\pi^{-}}|&=&1.17(26)~\mathrm{GeV},\nonumber\\
r_{f\pi K}&\equiv&\frac{|g_{f
K^+K^-}|}{|g_{f\pi^{+}\pi^{-}}|}=2.59(1.34)\ .
\end{eqnarray}
A large value of $r_{f\pi K}$ together with a narrow width disfavor a pure
$(u\bar{u}+d\bar{d})$ assignment of the $f_0(980)$, while its non-negligible width into $\pi\pi$
indicates that it cannot be a pure $s\bar{s}$ or
$K\bar{K}$ molecule.\\
-A possible gluonium component mixed with a $q\bar{q}$ state of the
$\sigma$ and $f_0(980)$ seems to be necessary for evading the
previous difficulties.
%%%%%%%%%%%%%%%%%%%%%%%%%%%%%%%%%%%%%%%%%%%%%%%%%%%%%%%%%%%%%%%%%%%%%%%%%%
\section{$\gamma\gamma\rightarrow\pi\pi$ process}
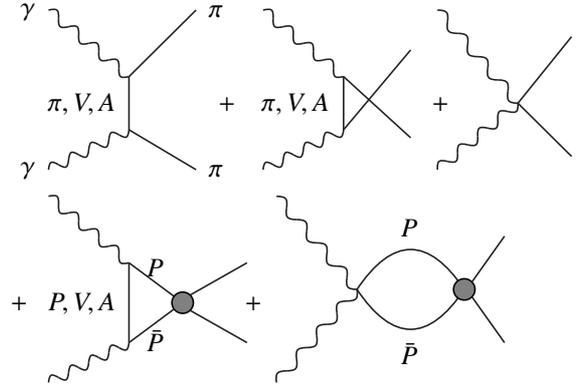
\begin{figure}
\begin{picture}(240,160)
  \SetWidth{0.6}
 % \Text(2,120)[l]{$T^{u}_{\pi}=$}
  \Text(20,145)[r]{$\gamma$}
  \Text(85,145)[l]{$\pi$}
  \Photon(25,145)(55,120){2}{4}
  \Photon(25,85)(55,100){2}{4}
  \Text(20,85)[r]{$\gamma$}
  \Text(85,85)[l]{$\pi$}
  \Line(55,120)(55,100)
  \Text(50,110)[r]{$\pi,V,A$}
  \Line(55,120)(80,145)
  \Line(55,100)(80,85)
  \Text(95,110)[r]{+}
  \Photon(105,145)(135,120){2}{4}
  \Photon(105,85)(135,100){2}{4}
  \Line(135,120)(135,100)
  \Text(130,110)[r]{$\pi,V,A$}
  \Line(135,120)(160,97)
  \Line(135,100)(160,130)
  \Text(175,110)[r]{+}
  \Photon(170,145)(200,110){2}{4}
  \Photon(170,85)(200,110){2}{4}
  \Line(200,110)(220,135)
  \Line(200,110)(220,90)
%  \Text(20,40)[l]{+}
  \Photon(25,75)(55,50){2}{4}
  \Photon(25,5)(55,20){2}{4}
  \Line(55,50)(55,20)
  \Text(50,35)[r]{$+~~~P,V,A$}
  \Line(55,50)(75,35)
  \Line(55,20)(75,35)
  \Text(65,45)[b]{$P$}
  \Text(65,25)[t]{$\bar{P}$}
  \GCirc(75,35){4}{0.5}
  \Line(78,38)(99,50)
  \Line(78,32)(99,20)
  \Text(105,35)[r]{$~~$+}
  \Photon(110,75)(140,40){2}{4}
  \Photon(110,5)(140,40){2}{4}
  \Curve{(140,40)(160,55)(180,40)}
  \Curve{(140,40)(160,25)(180,40)}
  \Text(160,60)[b]{$P$}
  \Text(160,20)[t]{$\bar{P}$}
  \GCirc(180,40){4}{0.5}
  \Line(183,43)(195,60)
  \Line(183,37)(195,20)
%  \Text(204,60)[l]{$(P=\pi,K)$}
%  \Text(204,45)[l]{R=V($1^{--}$),}
%  \Text(204,30)[l]{\ \ \ \ \ A($1^{++}$),}
%  \Text(204,15)[l]{\ \ \ \ \ B($1^{+-}$)}
\end{picture}
\caption{Born and Unitarized amplitudes $T^{u}_P$: $P\equiv
\pi,~K;~V\equiv \rho,~\omega;~A\equiv b_1, ~h_1,~a_1$.   }
\label{unitarized}
\end{figure}

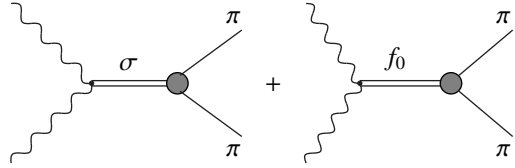
\begin{figure}
\begin{picture}(240,70)
%  \Text(10,40)[r]{$T^{D}_{\pi}=$}
  \Photon(20,65)(50,35){2}{4}
  \Photon(20,5)(50,35){2}{4}
  \Vertex(50,35){1}
  \Line(50,36)(79,36)
  \Line(50,34)(79,34)
  \Text(64,40)[b]{$\sigma$}
  \GCirc(82,35){4}{0.5}
  \Line(83,38)(106,55)
  \Line(83,32)(106,15)
  \Text(106,60)[r]{$\pi$}
  \Text(106,10)[r]{$\pi$}
  \Text(115,35)[l]{+}
  \Photon(130,65)(150,35){2}{4}
  \Photon(130,5)(150,35){2}{4}
  \Line(150,36)(180,36)
  \Line(150,34)(180,34)
  \Text(164,40)[b]{$f_0$}
  \Vertex(150,35){1}
  \GCirc(184,35){4}{0.5}
  \Line(187,38)(207,55)
  \Line(187,32)(207,15)
  \Text(207,60)[r]{$\pi$}
  \Text(207,10)[r]{$\pi$}
\end{picture}
\caption{Direct couplings of the resonances to $\gamma\gamma$}
\label{direct}
\end{figure}

\nin
The $I=0$ S-wave amplitude consists of two parts, the Born and unitarized
amplitudes shown in Fig.~\ref{unitarized}, and the direct $\gamma\gamma$
couplings shown in Fig.~\ref{direct}.
%%%%%%%%%%%%%%%%%%%%%%%%%%%%%%%%%%%%%%%%%%%%%%%%%%%%%%
%%%%%%%%%%%%%%%%%%%%%%%%%%%%%%%%%%%%%%%%%%%%%%%%%%%%%%%%%%%%%%
The unitarized amplitude can be calculated using chiral lagrangian.
The only ambiguity comes from the direct term. We introduce the direct
couplings of $\sigma$ and $f_0(980)$ to
$\gamma\gamma$~\cite{Mennessier83}:
\begin{equation}
T^{S}_{\pi}=\sqrt{2}\alpha
s[(f_{\sigma\gamma}+sf'_{\sigma\gamma})\tilde{T}_{\sigma\pi}+(f_{f_0\gamma}+sf'_{f_0\gamma})\tilde{T}_{f_0
\pi}]\ .
\end{equation}
for the S-wave and:
\begin{equation}
T^D_{\pi}=\frac{\alpha}{\sqrt{2}}[s^2 f^{\lambda=0}_{f_2\gamma}+s
f^{\lambda=2}_{f_2\gamma}]\tilde{T}_{f_2\pi}.
\end{equation}
for the D-wave with helicity 0 and 2 respectively. We refer to \cite{MNW1} for the expressions of the reduced amplitudes,
$\tilde{T}_{\sigma\pi}$, $\tilde{T}_{f_0\pi}$ and $\tilde{T}_{f_2\pi}$.\\
 We also include crossing channel contributions from
vector($1^{--}$) and axial-vector($1^{+-}$, $1^{++}$) exchanges
using normal vector
description.\\
For $I=0$ D-wave, we assume that it is dominated by the direct production of
$f_2(1270)$ with helicity 2~\cite{Close91}, and fix the direct
coupling of $f_2(1270)$ to $\gamma\gamma$ from the total width given by PDG,
$|f_{f_2\gamma}|=0.136\mathrm{GeV}^{-1}$. Indeed, we show in \cite{MNW1} that, at the $f_2$
mass, there is a cancellation between the unitarized and Born contributions. 
%\begin{figure}
%\begin{picture}(280,80)
% \SetWidth{0.6}
% \Photon(30,75)(60,40){2}{4}
%  \Photon(30,5)(60,40){2}{4}
%  \Text(25,75)[r]{$\gamma$}
%  \Text(25,5)[r]{$\gamma$}
%  \Vertex(60,40){1}
%  \Line(59,41)(89,41)
%  \Line(59,39)(89,39)
%  \Text(74,45)[b]{$f_2$}
%  \GCirc(92,40){4}{0.5}
%  \Line(93,44)(116,65)
%  \Line(93,36)(116,15)
%  \Text(116,70)[r]{$\pi$}
%  \Text(116,10)[r]{$\pi$}
%  \Text(130,40)[l]{$=\frac{\alpha}{\sqrt{2}} s f_{f_2\gamma} T_{\pi f_2}$}
%\end{picture}
%\caption{helicity 2 D-wave} \label{D2 wave}
%\end{figure}
%%%%%%%%%%%%%%%%%%%%%%%%%%%%%%%%%%%%%%%%%%%%%%%%%%%%%%%%%%%%%%%%%%%%%%%%%%%%%%%%%%%
\begin{figure}[hbt]
\centerline{\includegraphics[width=4.2cm]{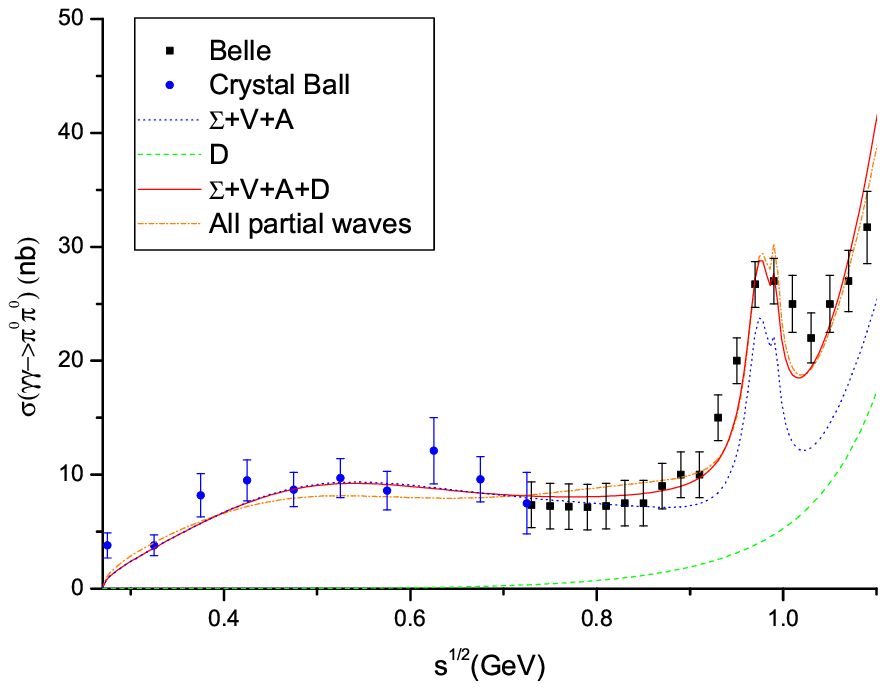}}
\centerline{\includegraphics[width=4.2cm]{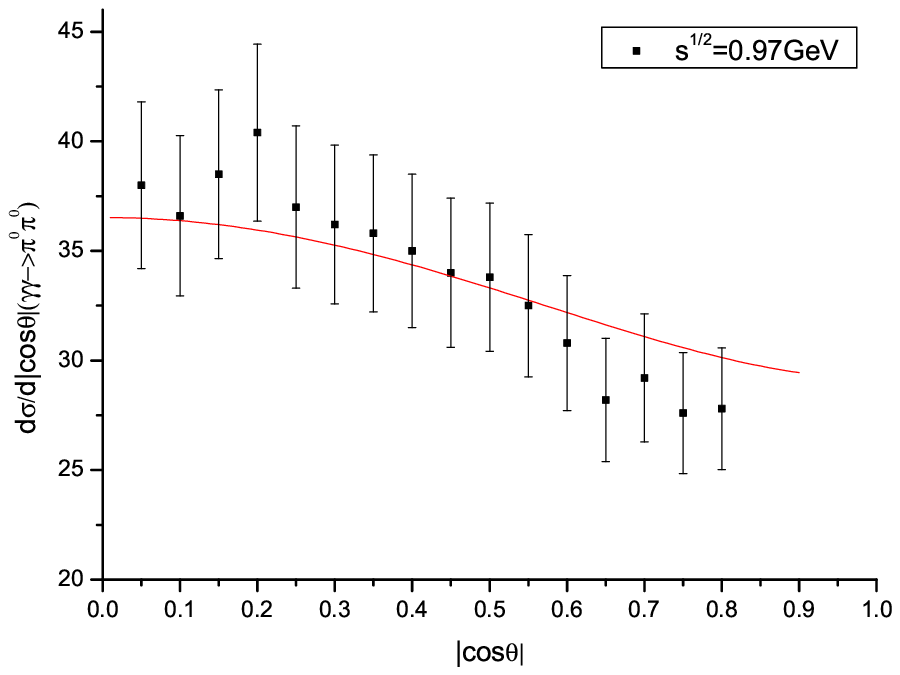}}
%{\epsfig{figure=mpsi2mc.eps,height=70mm}}
\caption{\scriptsize Fit up to $1.09\mathrm{GeV}$,
$\chi^2_{d.o.f}=\frac{39.5}{41}=0.96$. $\Sigma=\pi+K$. Dotted blue
(S-wave contribution); dashed green (D-wave contribution);
continuous red (S+D contribution); dash-dotted orange (all partial
wave contribution).} \label{fig6}
\end{figure}
\nin  The free parameters $f_{\sigma\gamma}$, $f'_{\sigma\gamma}$,
$f_{f_0\gamma}$ and $f'_{f_0\gamma}$ can be determined by fitting
the $\gamma\gamma\rightarrow\pi^0\pi^0$ total and differential cross
section data~\cite{CrystalBall}~\cite{Belle00}. The fit result is
shown in Fig.~\ref{fig6}. The $\gamma\gamma$ decay width from different
contributions are given in Tab.~\ref{tab:diphotonwidth}.
%%%%%%%%%%%%%%%%%%%%%%%%%%%%%%%%%%%%%%%%%%%%%%%%%%%%%%%
{\scriptsize
\begin{table}[hbt]
\setlength{\tabcolsep}{0.12pc}
 \caption{\scriptsize $\gamma\gamma$ decay width(in unit of $\mathrm{keV}$).
 $r_0\equiv\Gamma^{\lambda=0}_{f_2\rightarrow\gamma\gamma}/{\Gamma^{tot}_{f_2\rightarrow\gamma\gamma}}$}
    {\footnotesize
\begin{tabular}{ccccccccccc}\hline
\ & Set 2 & Set 3 & \cite{Ochs08} & \cite{Achasov08} &
\cite{Oller08} & \cite{Pennington08} & \cite{Pennington08} &
\cite{Zheng09} & \cite{Prades08} & PDG\cite{PDG08} \\ \hline
$\sqrt{s}$ & 1.09 & 1.09 & 0.8 & 1.44 & 0.8 & 1.44 & 1.44 & 1.4 & \
&
\ \\
$r_0$ & 0 & 0 & \  & \  & \  & 0.13 & 0.26 & 0.15 & \ & \ \\
\ & \ & \ & \ & \ & \ & \ & \ & \ & \ \\
$\Gamma^{dir}_{\sigma}$ & 0.16  & 0.20 & 0.13 & 0.01 & \ & \ & \ \\ % 0.06
$\Gamma^{resc}_{\sigma}$ & 1.53 & 1.43 & 2.70 & \ & \ & \ & \ & \ \\ % 2.4
$\Gamma^{tot}_{\sigma}$ & 3.11 & 3.10 & 3.90 & \ & 1.7 & 3.1 & 2.4 & 2.1 & 1.2 \\ % 1.8
\ & \ & \ & \ & \ & \ & \ & \ & \ & \ & \ \\
$\Gamma^{dir}_{f_0}$ & 0.29 & 0.27 & \ & 0.015 & \ & \ & \ & \ & \ & \ \\
$\Gamma^{resc}_{f_0}$ & 0.90 &  0.81 & \ & \ & \ & \ & \ & \ & \ & \ \\
$\Gamma^{tot}_{f_0}$ & 0.17 & 0.15 & \ & \ & \ & 0.42 & 0.10 &0.13 & \ & $0.29\pm0.08$ \\
%\ & \ & \ & \ & \ & \ \\
%\chi^2_{d.o.f} & $\frac{23.3}{24}=0.97$ & $\frac{21.6}{24}=0.90$ & \ & \ & \ \\
\hline
\end{tabular}
} \label{tab:diphotonwidth}
\end{table}
} \nin
%%%%%%%%%%%%%%%%%%%%%%%%%%%%%%%%%%%%%%%%%%%%
\section { Gluonium from $J/\psi,~\phi\to \gamma \sigma_B$ and $D_s\to l\nu \sigma_B$}
%%%%%%%%%%%%%%%%%%%%%%%%%%%%%%%%%%%%%%%%%%%%
\nin
The previous possible gluonium assignement of the $\sigma$ can be tested from $J/\psi,~\phi\to \gamma \sigma_B$ and $D_s\to l\nu \sigma_B$ processes ($\sigma_B$ is a n unmixed hypothetical gluonium state) as discussed respectively in \cite{VENEZIA,NSVZ,SNG0,SN06} and \cite{DOSCH}.
We expect to have the branching ratios $\times 10^3$:
\bea
&&B(J/\psi\to \sigma_B\gamma)\times B(\sigma_B\to {\rm all})\simeq  (0.4\sim 1.0)~,\nnb\\
&&B(\phi\to \sigma_B\gamma)\simeq 0.12~,
\eea
and:
\beq
{\Gamma[D_s\to\sigma_B(gg)l\nu]\over \Gamma[D_s\to S_2(\bar qq)l\nu]}\approx {1\over |f_+(0)|^2}\ga f_{\sigma_B}\over M_c\dr^2\simeq {\cal O}(1)~,
\eeq
for $M_c\approx 1.5$ GeV, $f_{\sigma_B}\approx 1$ GeV \cite{VENEZIA,SNG0,SN06}, where $|f_+(0)|\simeq 0.5$ \cite{SEMILEP} is the form factor associated to the $\bar qq$ semileptonic production. The rates of these productions are in fair agreement with existing data supporting again or not excluding a large gluonium component of the $\sigma$.
%%%%%%%%%%%%%%%%
\section{Conclusions}
\nin
%%%%%%%%%%%%%%%%
-We use an improved coupled channel K-matrix model, taking into
account Adler zero and left hand singularities. We extract the
pole positions and widths, as well as the hadronic and $\gamma\gamma$  couplings of
$\sigma$ and $f_0(980)$ by fitting experimental data.\\
-The values of their direct widths favour a large gluon content for
the $\sigma$ meson but are not decisive for explaining the
substructure of the $f_0(980)$ meson, which can mainly be either  a $\bar ss$ or a gluonium.\\
-The large values of the rescattering widths, due to meson loops,
can be also obtained if they are gluonia states but not necessarily
if they are four-quark or molecule states.
%%%%%%%%%%%%%%%%%%%%%%%%%%%
\section*{Acknowledgements}
\nin
%%%%%%%%%%%%%%%%
This work has been partly supported by CNRS-IN2P3 within the project Non-perturbative QCD and Hadron Physics. X.G. Wang thanks the Laboratoire de Physique Th\'eorique et Astroparticules (LPTA) of Montpellier for the hospitality. 



%% References


%\bibliographystyle{elsarticle-num}


 \end{document}